# A clinical trial engineering firm

Matthias Wjst[1]

[1]contact address

Institut für KI und Informatik in der Medizin

Lehrstuhl für Medizinische Informatik

TU München

Grillparzerstr. 18

81675 München

Germany

wjst@tum.de

http://orcid.org/0000-0002-4974-5631

Abstract

Paper mills produce fraudulent research manuscripts built on recycled tables and figures, or on entirely fabricated data. A more recent pattern has emerged: apparently genuine trials with real patients, but with manipulated statistical analyses engineered to support regulatory approval while remaining plausible to peer reviewers.

This analysis applies the INSPECT-SR trustworthiness framework to 23 randomised controlled trials and post-marketing studies linked to CinnaGen Co., Iran's largest biosimilar manufacturer, and its clinical operations subsidiary Orchid Pharmed. Papers were retrieved from PubMed and assessed against the original study records. A total of 180 problems were identified across nine categories. The five most frequent issues were reporting failures (n=37), arithmetic violations (n=28), design flaws (n=26), registration irregularities (n=25), and statistical errors (n=25).

Analysis of the co-authorship network shows that trial design, data management, and manuscript preparation were concentrated within the sponsoring organisation. The underlying structural drivers appear to be a convergence of domestic publication incentives, commercial pressure from international sanctions that created demand for domestically produced drugs, and regulatory pathways that require this body of trial evidence.

Because this pattern differs fundamentally from classical paper mills, we propose the term clinical trial engineering to describe it. Regulatory bodies, including the European Medicines Agency (EMA), should treat published clinical evidence from this cluster as unverified until independent access to individual participant data is granted.

Introduction

Paper mills are relatively well-defined operations, often commercial enterprises, that produce fraudulent scientific manuscripts for sale - typically to researchers who need publications for career advancement, institutional requirements, or grant eligibility (1, 2). Their methods vary. Some mills generate entirely synthetic data: tables and figures are produced from scratch, sometimes using templates recycled across multiple papers, which explains the tortured phrases and image duplication patterns found in forensic analyses. Others use real experimental infrastructure and produce genuine measurements, but with entirely fictitious authorship. A third variant harvests real data from public repositories or earlier papers and repackages it under new framing. What gets sold are authorship slots or ghostwritten papers built to a customer's specification. Detection is often straightforward: duplicated figures and tables across papers, implausible digit distributions, text plagiarism, and synonym substitution to evade similarity checks.

A distinct pattern has emerged more recently: apparently genuine trials with real patients, but with manipulated statistical analyses designed to support regulatory approval while remaining plausible to peer reviewers.

The commercial need for such studies does not arise in a vacuum. Western sanctions against Iran began after the 1979 US Embassy hostage crisis and expanded progressively, driven primarily by nuclear proliferation concerns, with the US, EU, and UN Security Council all participating. Their most acute pharmaceutical impact came in 2011-2012, when targeting of the Iranian banking system disrupted the financial infrastructure needed to import drugs and raw materials. Cancer and asthma medications showed the most severe shortfalls (3, 4). US pharmaceutical exports to Iran fell by half between 2011 and 2012, and Novartis reported that supply of life-saving medicines had been severely affected or fully halted, leaving an estimated six million Iranians with limited access to treatment for chronic diseases (3). At the crisis peak, over 350 medicines were in short supply, forcing the Iran Food and Drug Administration (IFDA) to implement emergency pricing and import policies to maintain even partial supply chains (4).

The sanctions regime therefore created both the political rationale and economic pressure for an aggressive domestic biosimilar program with regulatory fast-tracking of local equivalents as an explicit government priority. That structural context is what makes these studies not merely commercially viable, but nationally indispensable.

## Methods

The analysis presented here was initiated by the PREVENT-TAHA8 stem cell trial, published in the British Medical Journal in October 2025 and retracted five months later after many readers identified data concerns (5). Following my extended analysis of that paper and a letter to the BMJ (6), I received a message from an anonymous source indicating that a substantially larger body of questionable studies existed, connected to CinnaGen Co., an Iranian manufacturer of biosimilar drugs.

CinnaGen, founded in Tehran in 1994, is the largest biopharmaceutical manufacturer in the Middle East and North Africa (MENA) region, producing biosimilar and follow-on biological products across oncology, immunology, endocrinology, and rare disease. Orchid Pharmed is a wholly integrated subsidiary within the CinnaGen group that functions as the clinical operations arm - recruiting trial sites, managing data collection, drafting manuscripts, and holding the medical director function - appearing in the authorship and acknowledgement sections of every paper in this cluster as the entity that conducted, supervised, and wrote up the trials whose products CinnaGen manufactures and commercialises.

By late 2025, a new tool for assessing the trustworthiness of randomised controlled trials had become available. This approach was consecutively applied to 19 CinnaGen-affiliated studies of the past five years retrieved from PubMed. Results were posted to PubPeer. Not all anomalies indicate fraud; some may reflect unfamiliarity with RCT conventions, whether on my part or on that of the authors. Others, however, point toward systematic concealment or outright fabrication (7). An initial author network analysis suggested that CinnaGen itself may not be the primary driver of the cluster and indicated that further studies would be needed to characterise the role of Orchid Pharmed,. These additional studies were also submitted to PubPeer, although they could not convincingly separate both entities. Unfortunately, author responses to the PubPeer comments have been sparse, which is understandable given that hostilities between Israel and the United States on one side and Iran on the other escalated in February 2026.

The analysis presented here now covers 23 Iranian RCTs and post-marketing studies (7–30). ChatGPT 5.1 was used to expand the primary analysis of the published papers. Anthropic's Claude Sonnet 4.6 was used for classification of the findings. Only the second attempt seemed to classify correctly, as confirmed by manually checking a subset of the classifications.

The author network is constructed jointly for all CinnaGen and Orchid Pharmed PubMed entries by March 2026, with duplicates eliminated using the R library bibliometrix version 5.2.1 (31). The heatmap and Sankey plot were produced using D3.js (32).

Results

Errors were identified in all 23 examined papers. The 180 classified issues span nine categories, with reporting failures most frequent (n=37), followed by arithmetic violations (n=28) and statistical errors (n=25). Reporting failures encompassed CONSORT contradictions, safety table inconsistencies, and selective omission of pre-specified outcomes. Registration problems were near-universal: retrospective registration misrepresented as prospective, eligibility criteria differing between registry and publication, and undisclosed primary outcome changes. Statistical errors included competing-risks violations, inappropriate test statistics, and p-values irreconcilable with the reported group data.

Mapped to INSPECT-SR (33) (Table), domain D4 results accounted for most of the issues, led by statistical errors (check 4.9) and implausible outcome data (check 4.7). Domain D2 (governance and transparency) was dominated by methods plausibility (2.5) and registration timing (2.2). Domains D3 and D1 were minor contributors, the latter reflecting the absence of retractions or expressions of concern at the time of analysis. By predicted journal action, about one quarter may be classified as retraction-warranted. The retraction-predicted issues concentrate in four to five papers where numerous arithmetic impossibilities are not reconcilable with transcription or rounding error.

The main authorship network (Figure 3) resolves into a single connected component. HH (degree 13), the corporate chairperson of CinnaGen, is the highest-degree node - a role not conventionally associated with trial conduct or data analysis. HK (degree 8) and NA (degree 7) held the medical director function at the contracted trial operator sequentially. BL (degree 8) and MM (degree 8) are academic clinician and chemist at Tehran University of Medical

Sciences who bridge the university network to the corporate cluster. MM already carries 5 retractions and 44 PubPeer entries in adjacent fields.

Discussion

This analysis applied INSPECT-SR to clinical trial publications connected to a single Iranian biosimilar manufacturer and its clinical operations subsidiary and identified problems in every paper examined. The breadth of the findings - spanning reporting failures, arithmetic violations, design flaws, registration problems, and statistical errors across unrelated drug classes and a decade of publications - is inconsistent with isolated error or a single methodological weakness.

Two major uncertainties remain. First, the initial PubPeer reports have not been independently verified and may overlook important details or overstate minor findings. Second, the classification process is time-consuming and retains some subjectivity despite predefined criteria. This was the main reason for using a large language model to assist. That approach carries its own risks: in the initial run the model hallucinated classification rules, underscoring the need for supervision and validation. Minor discrepancies, however, are unlikely to alter the overall conclusions.

The published record cannot establish intent. Repeated arithmetically impossible findings - reported means falling below the lower bounds of their own confidence intervals, safety denominators exceeding the randomised sample size, standard deviations implying negative physiological values - are not reconcilable with transcription error or statistical inexperience. These concentrate in four to five papers that likely warrant retraction. Most issues, however, will probably trigger no journal action. Design-level failures such as absent non-inferiority frameworks, non-independent unit-of-analysis errors, and retrospective registrations misrepresented as prospective rarely justify retraction without regulatory engagement. The primary Iranian regulator showed no engagement when the BMJ editorial office approached it directly during the PREVENT-TAHA8 investigation (5). The cluster therefore reveals a structural gap in which documented data integrity failures fail to trigger enforcement absent a willing institutional actor.

The co-authorship structure supports this interpretation. Control of trial design, execution, data management, and manuscript preparation is concentrated within the sponsoring organisation, with academic co-authors providing institutional affiliation rather than independent oversight. Every actor benefits: the state demonstrates return on public investment, the manufacturer gains market access, the contract operator earns income, universities claim research leadership, journals collect fees, and authors accumulate reputation. The evidentiary record alone suffers.

The term "clinical trial engineering" is proposed here
to describe what this cluster represents - a form of research misconduct distinct from classical paper mills and considerably harder to detect. A clinical trial engineering operation works with real trials: the physicians are real, the substances tested are real, and the patients exist. What may be manipulated is the analysis, or simply the reporting of results. A clinical investigator who receives distorted results tables from an external statistical unit has no straightforward way to recognise the manipulation; they write a summary paper from numbers they cannot independently verify. There are no obviously invented data, and results typically appear plausible - or at least possible - to a superficial reviewer. What gets produced is not a publication sold to a customer, but something far more consequential: the evidential basis for a government to authorise market entry for a pharmaceutical product, and a public assurance that locally produced biosimilars are therapeutically equivalent to the imported originators they replace.

Detecting fraud in this context is an order of magnitude harder than in classical paper mills. It requires comparing trial registration documents against published methods sections and scrutinising baseline tables for even minor internal inconsistencies in reported statistics.

The underlying driver appears structural (34). Since around 2005, Iranian universities have linked faculty promotion, salary bonuses, and institutional rankings to indexed publication counts, including cash payments per paper at some institutions (35). This created demand for output regardless of quality. Notably, elevated error rates do not appear among Iranian scientists working abroad, ruling out cultural or individual explanations and pointing instead to domestic incentive structures.

International sanctions, as discussed in the introduction, compounded matters by restricting database access, conference participation, and international collaboration while simultaneously pressuring domestic drug production. The burden therefore does not rest on the Iranian system alone.

Two recommendations follow. First, regulatory bodies - including the European Medicines Agency (EMA) for pending or future CinnaGen submissions - should treat published clinical evidence from this cluster as unverified pending independent data access. The 2026 positive CHMP opinion for Zandoriah (teriparatide biosimilar), issued despite an existing PubPeer record (7), illustrates that post-publication integrity concerns are not yet reaching regulators - a gap that should be closed (36).

Second, and more fundamentally, access to individual participant data must precede regulatory reliance on trials with arithmetically doubtful summary statistics. Published records alone cannot confirm whether described patients were properly enrolled, treated, and followed. INSPECT-IPD (37), the extension of INSPECT-SR currently under development, addresses this by enabling verification at the level of individual patient data rather than aggregate statistics.

Table. Most frequently flagged INSPECT-SR checks across the 23-publication cluster, ranked by issue count. Check 4.9 (errors in statistical results) and check 4.7 (outcome data or effects implausible) are the most frequent findings.

| Checklist | Count(n) | Category |
|---|---|---|
| 4.9 | 31 | Errors in statistical results |
| 4.7 | 21 | Outcome data / effects implausible |
| 2.5 | 20 | Methods implausible given resources |
| 4.6 | 17 | Unexplained participant number inconsistencies |
| 2.2 | 13 | Registration timing/absence |
| 2.1 | 12 | Ethical approval concerns |
| 4.10 | 11 | Other contradictions in data |
| 2.3 | 10 | Publication vs. registration inconsistencies |
| 4.3 | 9 | Baseline data implausible |
| 4.5 | 8 | Loss-to-follow-up implausible |

Figure 1. Heatmap of 180 documented issues across 23 publications in the CinnaGen/Orchid Pharmed cluster, mapped to the 21 checks of the INSPECT-SR framework. Rows represent individual publications (labelled by drug or biological substance); columns represent INSPECT-SR checks grouped across four domains. Cell colour intensity reflects issue count, with grey indicating no issues identified. For details see the supplementals Excel file.

| | INSPECT-SR DOMAINS (Wilkinson 2025) | | | | | | | | | | | | | | | | | | | | | |
|---|---|---|---|---|---|---|---|---|---|---|---|---|---|---|---|---|---|---|---|---|---|---|
| | 1.1 | 1.2 | 1.3 | 2.1 | 2.2 | 2.3 | 2.4 | 2.5 | 3.1 | 3.2 | 4.1 | 4.2 | 4.3 | 4.4 | 4.5 | 4.6 | 4.7 | 4.8 | 4.9 | 4.10 | 4.11 | |
| | Retraction | Expression of concern | Other team studies raise concerns | Ethics approval concerns | Registration timing/ absence | Pub vs registration inconsistencies | Recruit-ment implausible | Methods implausible for resources | Duplicated/ incompatible content or text | Figure manip-ulation/ duplication | Eligibility criteria discre-pancies | Participant numbers implausible | Baseline data implausible | Figures/ tables/text discre-pancies | Follow-up numbers implausible | Participant number inconsis-tencies | Outcome data/effects implausible | Means/ variances impossible | Statistical errors | Other data contra-dictions | Cross-pub inconsis-tencies | INSPECT total |
| Omalizumab | | | | | 1 | | | 1 | | | | | | 1 | | 2 | | | 1 | 1 | | 7 |
| Ocrelizumab | | | | | | | | | | | | | 2 | 1 | 1 | | 1 | | | | | 5 |
| HPV vaccine (bivalent) | | | | 1 | 1 | | | 1 | | | | | | 1 | | 2 | 1 | | 2 | | 1 | 10 |
| Influenza vaccine (QIVr) | | | | | | | | | | | 1 | | | | | 1 | | 1 | 2 | | | 5 |
| Influenza vaccine (QIVr, baculovirus) | | | | 1 | | | 1 | 1 | | | | | | 1 | | | 1 | 1 | 1 | 2 | | 9 |
| Follitropin alfa | | | | | | | | 1 | | | | | 1 | 1 | 1 | 1 | | | 1 | | | 6 |
| Tacrolimus | | | | | | 4 | | | 1 | | | | | | | 1 | 1 | 2 | 2 | | | 11 |
| Laronidase | | | | | 1 | | 1 | 2 | | | | | | | 1 | 2 | 1 | | 1 | | | 9 |
| Somatropin | | | | 2 | 1 | | | 2 | 1 | | | | | | 2 | | | | | | | 8 |
| Factor VIII-Fc fusion (efmoroctocog α) | | | | 1 | 1 | | | 1 | | | | | | | | 2 | 4 | | 1 | | | 10 |
| Bevacizumab (biosimilar Stivant, PRP comparison) | | | | 1 | 1 | | | 1 | | 1 | | | | | | 1 | 2 | | 4 | | | 11 |
| Aflibercept | | | | 1 | 1 | | | | | | | | 1 | | | 3 | | | 2 | | 1 | 9 |
| Bevacizumab | | | | 1 | 1 | 2 | | 3 | | | | | | | 1 | | | | 1 | 1 | | 10 |
| Pertuzumab | | | | | 1 | | | | | | | | 1 | 1 | | | 1 | | 2 | | 1 | 7 |
| Interferon beta-1a (high vs low dose) | | | | 1 | | 1 | | | | | | | 1 | 1 | | 1 | | | 3 | | | 8 |
| Liraglutide | | | | 1 | | | 1 | 1 | | | 1 | | | | 1 | | | | | | | 5 |
| SARS-CoV-2 spike protein vaccine | | | | | | | | 2 | | | | | | | | | 3 | | | | 1 | 6 |
| Adalimumab | | | | | | | | | | | | | | | | | | | 2 | 1 | 1 | 4 |
| Trastuzumab (post-marketing) | | | | 1 | 1 | 1 | | 2 | | | | | | | | | 1 | | | 1 | | 7 |
| Denosumab | | 1 | | 1 | | 1 | | | | | | | 1 | | | | 1 | | 1 | 3 | 1 | 10 |
| Tacrolimus | | | | | 1 | | | 1 | 1 | | | | | | 1 | | | | 2 | | | 6 |
| Peginterferon beta-1a vs IFN-β1a | | | | | 1 | 1 | | | | | | | 1 | | | | 3 | 1 | 1 | | | 8 |
| Trastuzumab | | | | | 1 | | | 1 | | | | | 1 | | | 1 | 1 | | 2 | 2 | | 9 |
| | 0 | 1 | 0 | 12 | 13 | 10 | 3 | 20 | 3 | 1 | 2 | 0 | 9 | 7 | 8 | 17 | 21 | 5 | 31 | 11 | 6 | 180 |

Figure 2. Flowchart mapping 180 documented issues across 23 publications in the CinnaGen/Orchid Pharmed cluster to the INSPECT-SR framework. Flow proceeds left to right: study type, issue category, INSPECT-SR domain (D1–D4) and subtype. Band width is proportional to issue count.

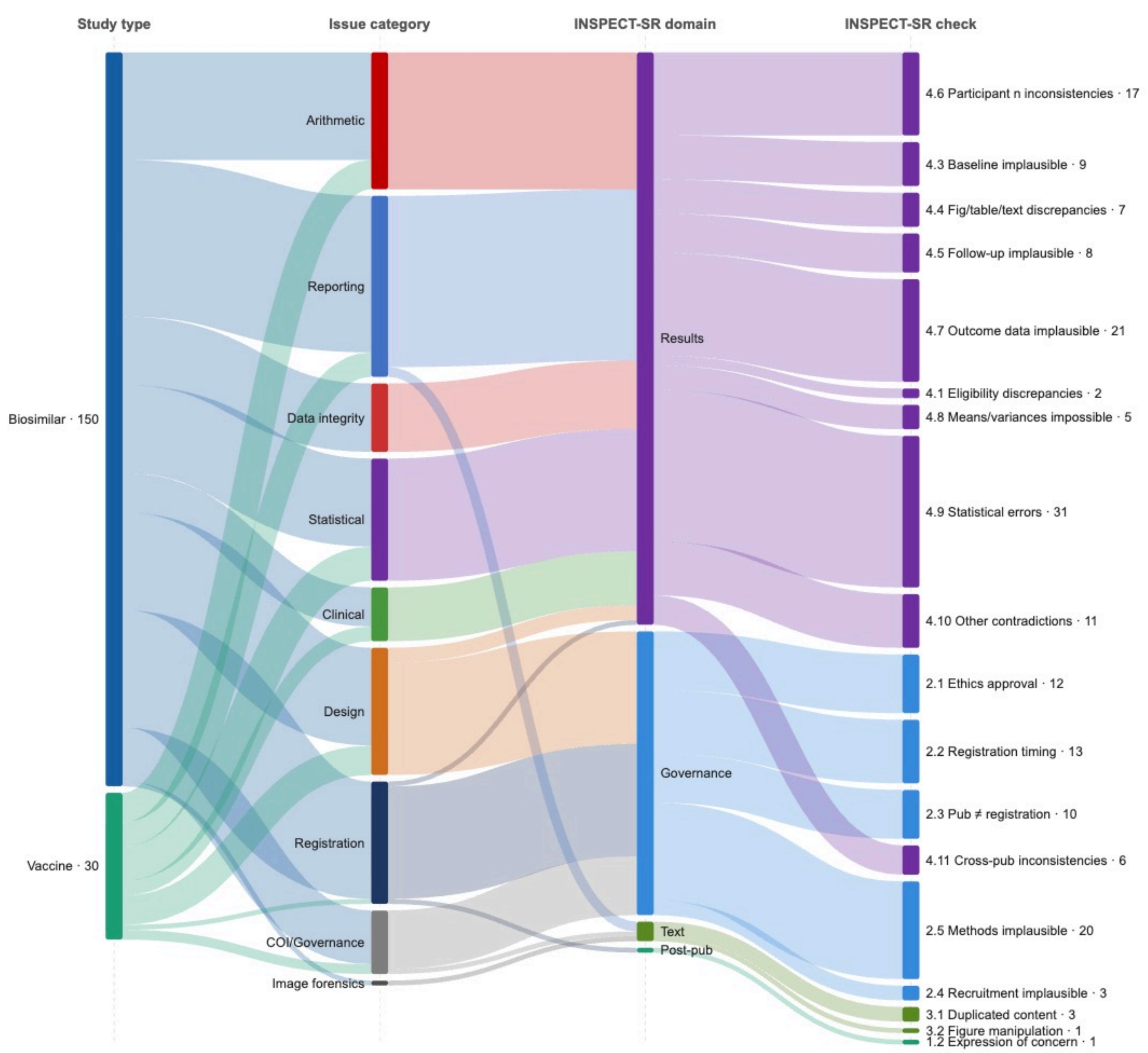

Figure 3. Co-authorship network of the CinnaGen/Orchid Pharmed cluster (top n = 50 authors). The peripheral cloud represents clinical trial site investigators. The red cluster contains the Orchid Pharmed authors HK (Medical Director) and NA (Head of Medical Affairs), together with AS (role unidentified). The second cluster includes MM (medicinal chemist, Tehran University of Medical Sciences; subject of numerous retractions), HH (CEO, CinnaGen), and BL (former Chancellor, Tehran University of Medical Sciences; brother of the President of the Iranian Judiciary).

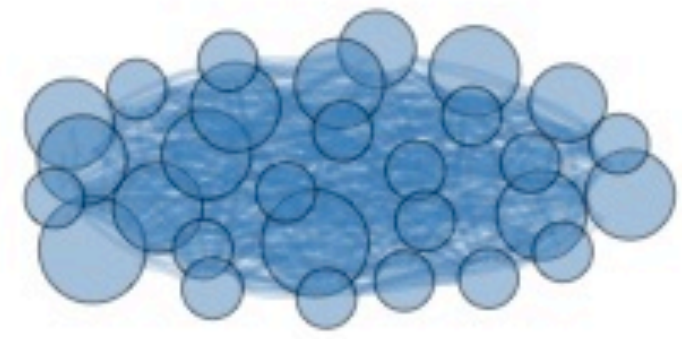

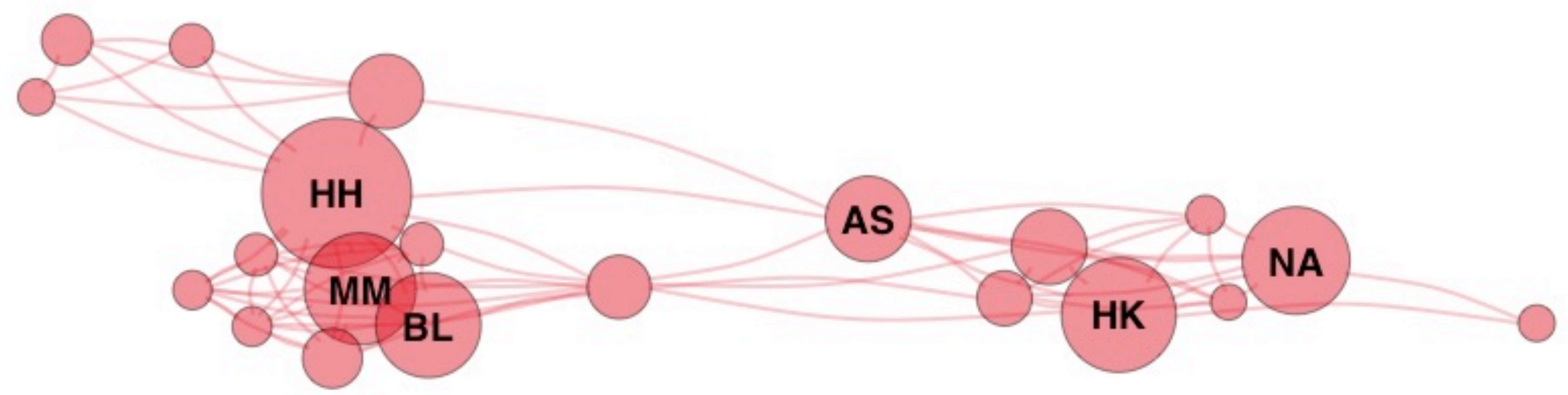

Funding

None.

Ethics / Data availability

Only public available data have been used where all sources are reported in the paper.

Conflicts of Interest

The author declares no conflicts of interest.

Views expressed in this analysis are solely those of the author and do not necessarily reflect the positions of any affiliated institutions.

Author contribution

Study design, analysis and writing by the author.